\documentclass[aps,pra,floatfix,showpacs,10pt,twocolumn]{revtex4}
\usepackage[dvips]{graphicx}
\usepackage{amsmath,amsfonts,amssymb,graphics,graphicx,epsfig,color,times,bbm}

\begin{document}

\title{Storage and retrieval of continuous-variable polarization-entangled cluster states in atomic ensembles}
\author{Da-Chuang Li$^{1,2}$\thanks{%
dachuangli@ustc.edu.cn}, Chun-Hua Yuan$^{1}$, Zhuo-Liang
Cao$^{2}$, Weiping Zhang$^{1}$\thanks{%
wpzhang@phy.ecnu.edu.cn (Corresponding Author)}}
\affiliation{$^{1}$Quantum Institute for Light and Atoms, Department of Physics, East
China Normal University, Shanghai 200062, People's Republic of China\\
$^{2}$Department of Physics and Electronic Engineering, Hefei Normal
University, Hefei 230601, People's Republic of China}
\date{\today }

\begin{abstract}
We present a proposal for storing and retrieving a
continuous-variable quadripartite polarization-entangled cluster
state, using macroscopic atomic ensembles in a magnetic field. The
Larmor precession of the atomic spins leads to a symmetry between
the atomic canonical operators. In this scheme, each of the four
spatially separated pulses passes twice through the respective
ensemble in order to map the polarization-entangled cluster state
onto the long-lived atomic ensembles. The stored state can then be
retrieved by another four read-out pulses, each crossing the
respective ensemble twice. By calculating the variances, we analyzed
the fidelities of the storage and retrieval, and our scheme is
feasible under realistic experimental conditions.
\end{abstract}

\maketitle

\section{Introduction}

In recent years, the investigation of continuous-variable (CV)
quantum information has attracted much interest due to the relative
simplicity and high efficiency in the generation, manipulation, and
detection of the CV quantum state \cite{Braunstein05,Hammerer}. As
important resources, the CV entangled states have been widely
applied in various quantum information processes, such as quantum
teleportation \cite{Braunstein98}, dense coding \cite{Jing03},
entanglement swapping \cite{Polkinghorne99}, quantum telecloning
\cite{Loock01}, and quantum computation \cite{Lloyd99,Gu09}.
Quite recently, Zhang and Braunstein introduced CV cluster states \cite%
{Zhang06}, which are different from CV Greenberger-Horne-Zeilinger
(GHZ) states, and the entanglement of the states is harder to
destroy than that of GHZ states. In one-way quantum computation, CV
cluster states play an important role \cite{Gu09}, and all multimode
Gaussian operations, performed through the cluster states, need only
homodyne detection \cite{Menicucci06}.

In view of the important functions of CV cluster states in quantum
information processing, as well as the storage and retrieval of
quantum states, we presented a proposal for the quantum memory of
discrete-variable cluster states \cite{Yuan}. Furthermore, it is
also worth initiating a study on the reversible memory of CV cluster
states. In addition, compared to CV quadrature entanglement
(corresponding to the amplitude and phase quadratures), the CV
polarization entanglement (corresponding to the polarization basis)
introduced by Korolkova \emph{et al.} \cite{Korolkova02} has the
advantages of compatibility with the spin variables of an atomic
system \cite{Cerf07} and of direct detection of its Stokes operators
(not requiring complicated local oscillator measurements)
\cite{Bowen88,Bowen89,Dong07}. Therefore, it is necessary to
investigate the storage and retrieval of CV polarization-entangled
cluster states (PECSs), and this paper is also an extension of our
memory scheme for discrete-variable cluster states \cite{Yuan}.

In this paper, we propose a protocol for efficiently storing and
retrieving the CV PECSs in atomic ensembles. For quantum memory, one
distinct method is based on the Faraday effect \cite{Kuzmich03}.
Using this approach, only partial transfer of the quantum
information can be realized based on the quantum nondemolition (QND)
interaction, but this shortcoming can be overcome by additional
measurement and feedback \cite{Julsgaard04} or by use of a
multipassage geometric configuration
\cite{Takano08,Sherson06,Fiur06}. The most efficient schemes for
this approach are eight passages of a single pulse \cite{Sherson06}
or two pulses each passing twice \cite{Fiur06}. To reduce the
complexity, Muschik \emph{et al.} recently proposed a simple
scenario \cite{Muschik06} which can realize the complete transfer of
a quantum state between light and an atomic ensemble in a magnetic
field. Here, our scheme is also partly an extension of the existing
work \cite{Muschik06}. In our scheme, the storage of the
quadripartite CV PECS can be achieved by using four spatially
separated pulses (write-in pulses), which are initially in the CV
PECS. Based on the Larmor precession of atoms in a magnetic field,
each of the pulses passes through the corresponding atomic ensemble
twice. After the storage time, the read-out pulses (another four
pulses) are sent through the ensembles twice just like the write-in
pulses. In this step, the CV PECS that is stored in the atomic
ensembles will be transferred to the read-out pulses. Thus, we
realize the storage and retrieval of the CV PECS, which approaches
perfection with increasing coupling strength of the light fields
interacting with the atomic ensembles. Moreover, we discuss the
feasibility of our scheme, which can be realized under realistic
experimental conditions.

This paper is organized as follows. In Sec. II, we propose a
concrete scheme for storing and retrieving the CV PECS in atomic
ensembles. In Sec. III, we analyze the fidelities of the storage and
retrieval by calculating the variances. In Sec. IV, we discuss the
experimental feasibility of the current scheme. Finally, in Sec. V,
a conclusion is given.

\section{Storage and Retrieval of the CV PECS}

CV cluster-type (graph-type) states are defined as states that
become zero eigenstates of a set of quadrature combinations in the
limit of infinite squeezing \cite{Zhang06,Loock07},
\begin{equation}
\Big(\hat{p}_{a}-\sum_{\substack{ b\in N_{a}}}\hat{x}_{b}\Big)\rightarrow 0,%
\text{ \ }\forall a\in G.  \label{1}
\end{equation}%
Here, the dimensionless amplitude and phase operators $\hat{x}_{a}$ and $%
\hat{p}_{a}$, which satisfy the commutation relation $[\hat{x}_{a},\hat{p}%
_{a}]=i/2$, correspond to the quadratures of an optical mode $a$ with
annihilation operator $\hat{a}_{a}=\hat{x}_{a}+i\hat{p}_{a}$. Each mode $%
a\in G$ corresponds to a vertex of the graph $G$ while the modes
$b\in N_{a} $ are the nearest neighbors of mode $a$. For the
quadripartite CV cluster states, there are three different kinds of
structures, i.e., the linear cluster state, the square cluster
state, and the T-shaped cluster state \cite{Loock07,Yukawa08}. Here
we use the linear cluster state as an example to describe the memory
of the CV PECS whose corresponding operators can be expressed as
\cite{Yukawa08}
\begin{equation}
\begin{split}
\hat{X}_{L_{1}}& =\frac{1}{\sqrt{2}}e^{r_{1}}\hat{X}_{L_{1}}^{(0)}+\frac{1}{%
\sqrt{10}}e^{r_{2}}\hat{X}_{L_{2}}^{(0)}-\frac{2}{\sqrt{10}}e^{-r_{3}}\hat{P}%
_{L_{3}}^{(0)}, \\
\hat{P}_{L_{1}}& =\frac{1}{\sqrt{2}}e^{-r_{1}}\hat{P}_{L_{1}}^{(0)}+\frac{1}{%
\sqrt{10}}e^{-r_{2}}\hat{P}_{L_{2}}^{(0)}+\frac{2}{\sqrt{10}}e^{r_{3}}\hat{X}%
_{L_{3}}^{(0)}, \\
\hat{X}_{L_{2}}& =-\frac{1}{\sqrt{2}}e^{-r_{1}}\hat{P}_{L_{1}}^{(0)}+\frac{1%
}{\sqrt{10}}e^{-r_{2}}\hat{P}_{L_{2}}^{(0)}+\frac{2}{\sqrt{10}}e^{r_{3}}\hat{%
X}_{L_{3}}^{(0)}, \\
\hat{P}_{L_{2}}& =\frac{1}{\sqrt{2}}e^{r_{1}}\hat{X}_{L_{1}}^{(0)}-\frac{1}{%
\sqrt{10}}e^{r_{2}}\hat{X}_{L_{2}}^{(0)}+\frac{2}{\sqrt{10}}e^{-r_{3}}\hat{P}%
_{L_{3}}^{(0)}, \\
\hat{X}_{L_{3}}& =-\frac{2}{\sqrt{10}}e^{r_{2}}\hat{X}_{L_{2}}^{(0)}-\frac{1%
}{\sqrt{10}}e^{-r_{3}}\hat{P}_{L_{3}}^{(0)}-\frac{1}{\sqrt{2}}e^{-r_{4}}\hat{%
P}_{L_{4}}^{(0)}, \\
\hat{P}_{L_{3}}& =-\frac{2}{\sqrt{10}}e^{-r_{2}}\hat{P}_{L_{2}}^{(0)}+\frac{1%
}{\sqrt{10}}e^{r_{3}}\hat{X}_{L_{3}}^{(0)}+\frac{1}{\sqrt{2}}e^{r_{4}}\hat{X}%
_{L_{4}}^{(0)}, \\
\hat{X}_{L_{4}}& =\frac{2}{\sqrt{10}}e^{-r_{2}}\hat{P}_{L_{2}}^{(0)}-\frac{1%
}{\sqrt{10}}e^{r_{3}}\hat{X}_{L_{3}}^{(0)}+\frac{1}{\sqrt{2}}e^{r_{4}}\hat{X}%
_{L_{4}}^{(0)}, \\
\hat{P}_{L_{4}}& =-\frac{2}{\sqrt{10}}e^{r_{2}}\hat{X}_{L_{2}}^{(0)}-\frac{1%
}{\sqrt{10}}e^{-r_{3}}\hat{P}_{L_{3}}^{(0)}+\frac{1}{\sqrt{2}}e^{-r_{4}}\hat{%
P}_{L_{4}}^{(0)}.
\end{split}
\label{cluster state}
\end{equation}%
In this paper, the operators $\hat{X}_{L_{i}}$ and
$\hat{P}_{L_{i}}$\ are the light polarization canonical variables
where the superscript (0) denotes the initial condition and
$r_{i}(i=1,2,3,4)$ is the polarization squeezing parameter of the
$i$th \textquotedblleft mode".

\begin{figure}[tbp]
\includegraphics[scale=0.45,angle=0,bb=53 200 543 629]{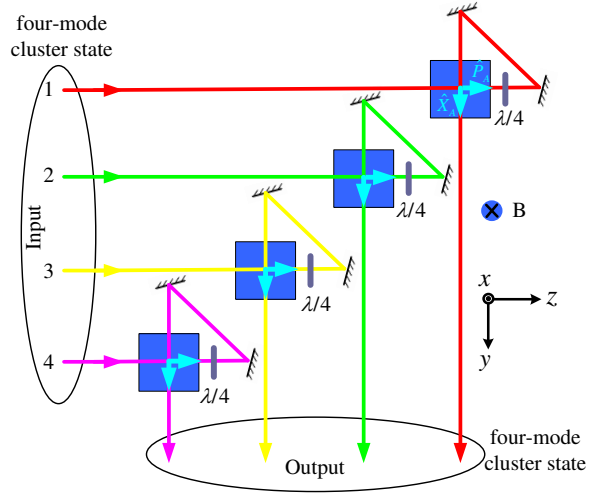}
\caption{(Color online) Schematic for storage or retrieval of quadripartite
CV cluster states. Here, four optical pulses (four-mode cluster state) each
propagates through the corresponding x-polarized atomic spin ensemble twice.
For each pulse, the $\hat{V}_{1}$ interaction happens during the first
passage (z direction), then each pulse is sent through a $\protect\lambda /4$
wave plate, and is reflected back onto the atomic ensemble, thus the $\hat{V}%
_{2}$ interaction occurs during the second passage (y direction).}
\label{fig1}
\end{figure}

\begin{figure}[tbp]
\includegraphics[scale=1,angle=0,bb=128 493 333 648]{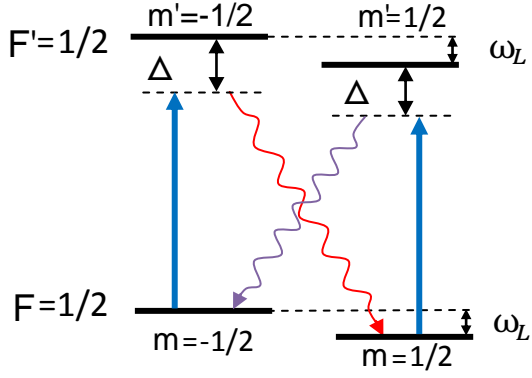}
\caption{(Color online) Level structure of the atoms. The classical field
(thick lines) drives the $m=\pm1/2\rightarrow m^{\prime }=\pm1/2$
transitions, and the co-propagating quantum fields (wavy lines) couples to $%
m=\mp1/2\rightarrow m^{\prime }=\pm1/2$. $\protect\omega_{L}$ is the Larmor
frequency, and $\Delta$ is the detuning.}
\label{fig2}
\end{figure}

The schematic diagram of the proposed experimental system is shown
in Fig. \ref{fig1} where four atomic ensembles spin polarized along
$x$ are placed in the magnetic field and used as memory cells. The
relevant level structure of the atoms is shown in Fig. \ref{fig2}.
These four ensembles form four equivalent channels, each of which is
used to store the corresponding polarization-squeezed light field.
Using a Holstein-Primakoff transformation
and approximation, the atomic canonical operators $\hat{X}_{A_{i}}$ and $%
\hat{P}_{A_{i}}$ (the subscript $A$ denotes the atomic ensemble) can be
defined as the $y$ and $z$ components of the collective angular momentum $%
\hat{J}$, i.e., $\hat{X}_{A_{i}}=(\hat{J}_{y})_{i}/\sqrt{\langle (\hat{J}%
_{x})_{i}\rangle }$ and $\hat{P}_{A_{i}}=(\hat{J}_{z})_{i}/\sqrt{\langle (%
\hat{J}_{x})_{i}\rangle }$ where $i=1,2,3,4$. The Larmor precession
of the atomic spins in the magnetic field is a characteristic
feature of this
protocol; it leads to a symmetry between the operators $\hat{X}_{A}$ and $%
\hat{P}_{A}$. Four spatially separated optical pulses propagate
along $z$ where each pulse consists of a strong $x$-polarized
component and a copropagating quantum field with $y$ polarization.
The classical light field drives the $m=\pm 1/2\rightarrow m^{\prime
}=\pm 1/2$ transitions while the copropagating quantum field couples
to $m=\mp 1/2\rightarrow m^{\prime
}=\pm 1/2$. For input light fields, the Stokes vector component $(\hat{S}%
_{x})_{i}$ is a macroscopic classical quantity and we can define the
operators $\hat{X}_{L_{i}}=(\hat{S}_{y})_{i}/\sqrt{\langle (\hat{S}%
_{x})_{i}\rangle }$ and $\hat{P}_{L_{i}}=(\hat{S}_{z})_{i}/\sqrt{\langle (%
\hat{S}_{x})_{i}\rangle }$ where $i=1,2,3,4$. According to Eq.
(\ref{1}), the
four light pulses are the $P$-squeezed states, i.e., the variance of the $\hat{S}%
_{z}$ Stokes operator is less than the coherent state value \cite%
{Korolkova02}.

In order to explain our scheme better, we first consider one pulse
of four spatially separated pulses. When a pulse passes through the
cubic atomic ensemble twice as depicted in Fig. \ref{fig1}, the
Hamiltonian of the interaction can be written as
\cite{Muschik06,Echaniz}
\begin{eqnarray}
\hat{H} &=&\hat{H}_{A}+\hat{H}_{L}+\hat{V}_{1}+\hat{V}_{2}, \\
\hat{V}_{1} &=&\frac{\hbar \kappa }{\sqrt{T}}\hat{P}_{A}\hat{P}_{L}(0),
\notag \\
\hat{V}_{2} &=&\frac{\hbar \kappa }{\sqrt{T}}\hat{X}_{A}\hat{X}_{L}(d),
\notag
\end{eqnarray}%
where $\hat{H}_{A}$ represents the Zeeman splitting of the atomic ground
state, causing Larmor precession of the transverse spin components $\hat{X}%
_{A}$ and $\hat{P}_{A}$, and $\hat{H}_{L}$ describes the free
propagation of light. The two left terms $\hat{V}_{1}$ and
$\hat{V}_{2}$ denote the off-resonant scattering interaction in the
first and the second passage of the pulse respectively, $\kappa $ is
the coupling strength, and $T$ is the duration of the pulse. The
argument \textquotedblleft $0$\textquotedblright\ of the light
operator $\hat{P}_{L}$ in $\hat{V}_{1}$ indicates that the first
scattering interaction occurs at $r=0$. The argument
\textquotedblleft $d$\textquotedblright\ of the light operator
$\hat{X}_{L}$ in $\hat{V}_{2}$ means that the second scattering
interaction happens after the light has traveled some distance $d$
in the small loop between the mirrors. After the first passage, the
pulse is sent through a quarter-wave plate, which interchanges the
light operators $\hat{P}_{L}$ and $\hat{X}_{L}$.
Furthermore, the changed geometry also interchanges the atomic operators $%
\hat{P}_{A}$ and $\hat{X}_{A}$.

Now, we explain the storage and retrieval of the four-mode CV PECS.
For the procedure of storage, after the four
pulses pass twice through the four corresponding atomic ensembles (see Fig. %
\ref{fig1}), whose operators $\hat{X}_{L_{i}}^{in}$ and
$\hat{P}_{L_{i}}^{in} $ ($i=1,2,3,4$) have the forms of Eq.
(\ref{cluster state}), the output of the atomic ensembles will have
the following relations with the input of the atomic ensembles and
the light pulses \cite{Muschik06}:
\begin{equation}
\left(
\begin{array}{c}
\hat{X}_{A_{1}}^{out} \\
\hat{P}_{A_{1}}^{out} \\
\hat{X}_{A_{2}}^{out} \\
\hat{P}_{A_{2}}^{out} \\
\hat{X}_{A_{3}}^{out} \\
\hat{P}_{A_{3}}^{out} \\
\hat{X}_{A_{4}}^{out} \\
\hat{P}_{A_{4}}^{out}%
\end{array}%
\right) =e^{-\kappa ^{2}/2}\left(
\begin{array}{c}
\hat{X}_{A_{1}}^{in} \\
\hat{P}_{A_{1}}^{in} \\
\hat{X}_{A_{2}}^{in} \\
\hat{P}_{A_{2}}^{in} \\
\hat{X}_{A_{3}}^{in} \\
\hat{P}_{A_{3}}^{in} \\
\hat{X}_{A_{4}}^{in} \\
\hat{P}_{A_{4}}^{in}%
\end{array}%
\right) +\sqrt{1-e^{-\kappa ^{2}}}\left(
\begin{array}{c}
\hat{X}_{L_{1}}^{in} \\
\hat{P}_{L_{1}}^{in} \\
\hat{X}_{L_{2}}^{in} \\
\hat{P}_{L_{2}}^{in} \\
\hat{X}_{L_{3}}^{in} \\
\hat{P}_{L_{3}}^{in} \\
\hat{X}_{L_{4}}^{in} \\
\hat{P}_{L_{4}}^{in}%
\end{array}%
\right) .  \label{input}
\end{equation}%
Obviously, for large values of $\kappa $, $\hat{X}_{L_{i}}^{in}\rightarrow \hat{X}%
_{A_{i}}^{out}$ and $\hat{P}_{L_{i}}^{in}\rightarrow \hat{P}_{A_{i}}^{out}$ (%
$i=1,2,3,4$); thus, the output canonical operators of atomic
ensembles satisfy the following correlations:
\begin{equation}
\begin{split}
\hat{P}_{A_{1}}^{out}-\hat{X}_{A_{2}}^{out}& =\sqrt{2}e^{-r_{1}}\hat{P}%
_{A_{1}}^{(0)}, \\
\hat{P}_{A_{2}}^{out}-\hat{X}_{A_{1}}^{out}-\hat{X}_{A_{3}}^{out}& =\frac{%
\sqrt{10}}{2}e^{-r_{3}}\hat{P}_{A_{3}}^{(0)}+\frac{1}{\sqrt{2}}e^{-r_{4}}%
\hat{P}_{A_{4}}^{(0)}, \\
\hat{P}_{A_{3}}^{out}-\hat{X}_{A_{2}}^{out}-\hat{X}_{A_{4}}^{out}& =\frac{1}{%
\sqrt{2}}e^{-r_{1}}\hat{P}_{A_{1}}^{(0)}-\frac{\sqrt{10}}{2}e^{-r_{2}}\hat{P}%
_{A_{2}}^{(0)}, \\
\hat{P}_{A_{4}}^{out}-\hat{X}_{A_{3}}^{out}& =\sqrt{2}e^{-r_{4}}\hat{P}%
_{A_{4}}^{(0)}.
\end{split}
\label{correlation}
\end{equation}%
Thus, the storage is realized perfectly.

For the procedure of retrieval, another four read-out pulses are sent
through the four corresponding atomic ensembles twice with the same
interaction happening. Considering the contribution of the atomic ensembles
and the read-out pulses, the output can be written as \cite{Muschik06}
\begin{equation}
\left(
\begin{array}{c}
\hat{X}_{L_{1}}^{^{\prime }out} \\
\hat{P}_{L_{1}}^{^{\prime }out} \\
\hat{X}_{L_{2}}^{^{\prime }out} \\
\hat{P}_{L_{2}}^{^{\prime }out} \\
\hat{X}_{L_{3}}^{^{\prime }out} \\
\hat{P}_{L_{3}}^{^{\prime }out} \\
\hat{X}_{L_{4}}^{^{\prime }out} \\
\hat{P}_{L_{4}}^{^{\prime }out}%
\end{array}%
\right) =-\sqrt{1-e^{-\kappa ^{2}}}\left(
\begin{array}{c}
\hat{X}_{A_{1}}^{^{\prime }in} \\
\hat{P}_{A_{1}}^{^{\prime }in} \\
\hat{X}_{A_{2}}^{^{\prime }in} \\
\hat{P}_{A_{2}}^{^{\prime }in} \\
\hat{X}_{A_{3}}^{^{\prime }in} \\
\hat{P}_{A_{3}}^{^{\prime }in} \\
\hat{X}_{A_{4}}^{^{\prime }in} \\
\hat{P}_{A_{4}}^{^{\prime }in}%
\end{array}%
\right) +e^{-\kappa ^{2}/2}\left(
\begin{array}{c}
\hat{X}_{L_{1}}^{^{\prime }in} \\
\hat{P}_{L_{1}}^{^{\prime }in} \\
\hat{X}_{L_{2}}^{^{\prime }in} \\
\hat{P}_{L_{2}}^{^{\prime }in} \\
\hat{X}_{L_{3}}^{^{\prime }in} \\
\hat{P}_{L_{3}}^{^{\prime }in} \\
\hat{X}_{L_{4}}^{^{\prime }in} \\
\hat{P}_{L_{4}}^{^{\prime }in}%
\end{array}%
\right) ,  \label{output}
\end{equation}%
where the prime is used to distinguish the retrieval process from the
storage process.

\begin{figure}[tbp]
\includegraphics[scale=0.55,angle=0,bb=90 263 491 566]{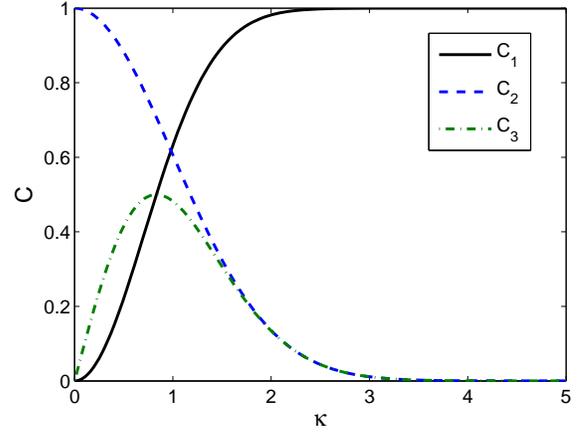}
\caption{(Color online) The coefficients versus coupling strength $\protect%
\kappa$. $C_{1}$, $C_{2}$ and $C_{3}$ are the coefficients of Eq. (\protect
\ref{coefficient}).}
\label{fig3}
\end{figure}

\begin{figure*}[tbp]
\includegraphics[scale=0.55,angle=0,bb=90 256 516 563]{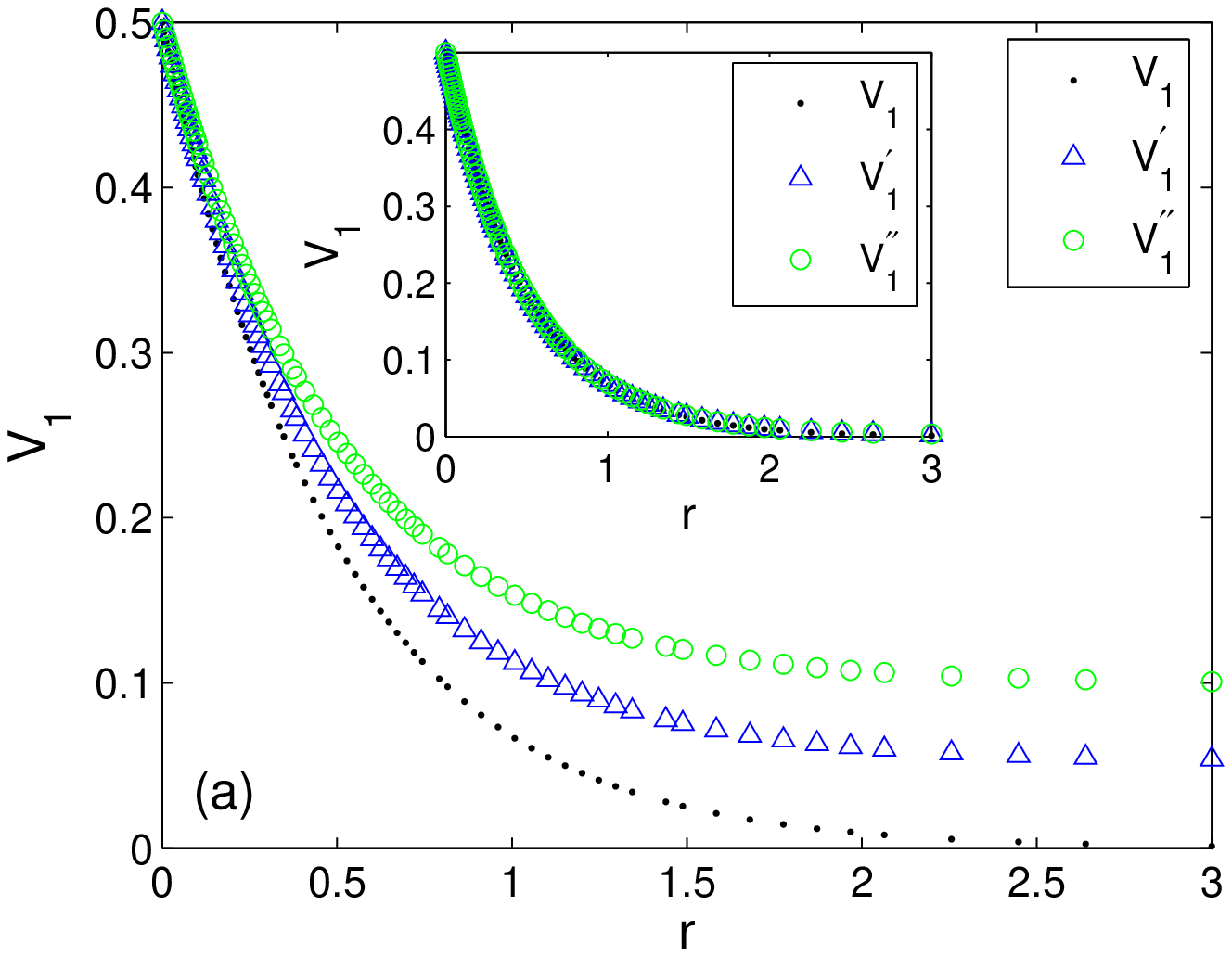} %
\includegraphics[scale=0.55,angle=0,bb=90 256 473 565]{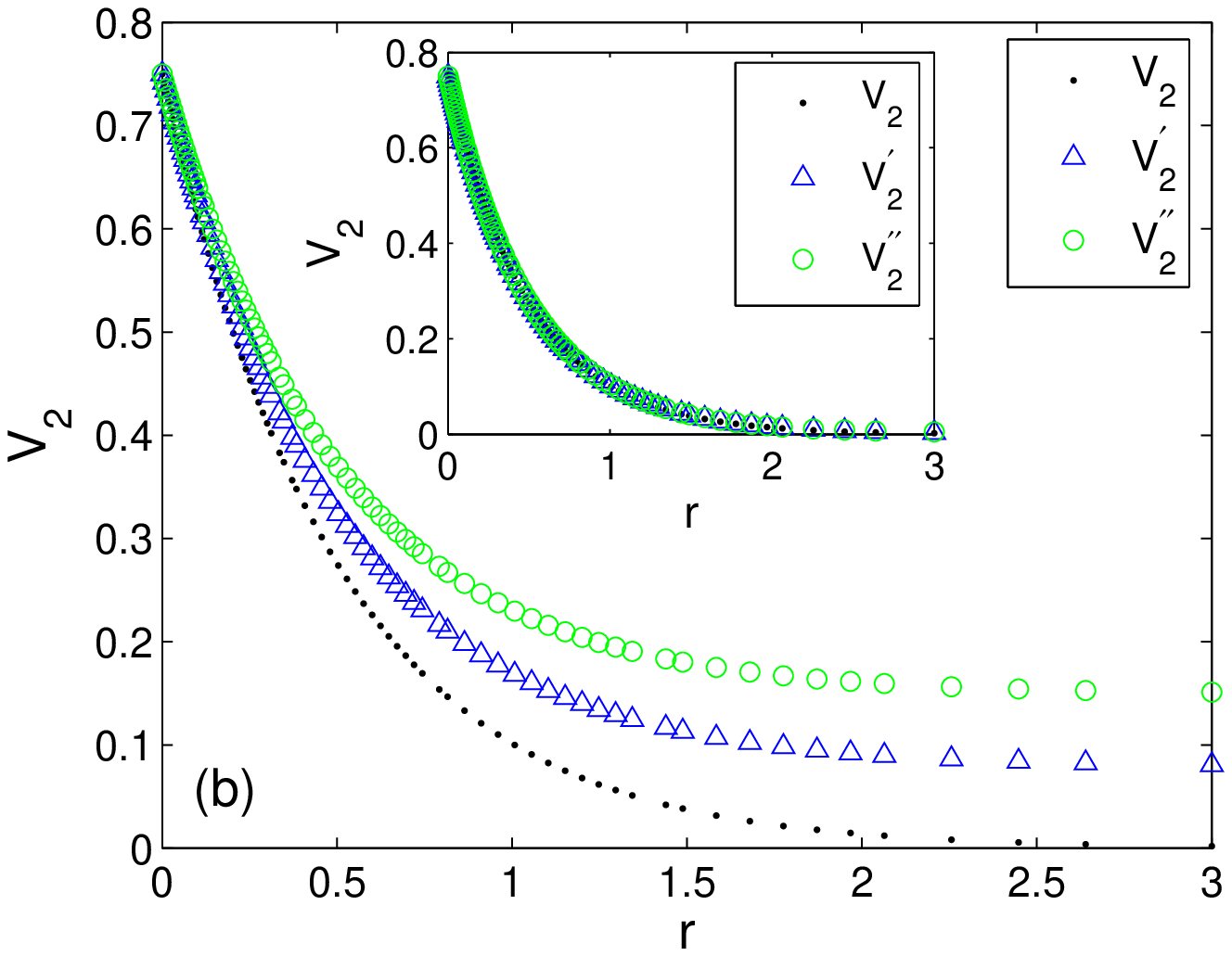}
\caption{(Color online) (a) The variances $V_{1}$, $V_{1}^{^{\prime }}$ and $%
V_{1}^{^{\prime \prime }}$ versus the squeezing parameter $r$. (b) The
variances $V_{2}$, $V_{2}^{^{\prime }}$ and $V_{2}^{^{\prime \prime }}$
versus the squeezing parameter $r$. Here, the big graphs and the inserted
small graphs correspond to $\protect\kappa =1.5$ and $\protect\kappa =2.5$,
respectively. }
\label{fig4}
\end{figure*}

After the whole storage and subsequent retrieval procedure we
consider the final output. For atomic ensembles, the output
operators of the write-in
procedure are the input variables of the read-out procedure, i.e., $\hat{X}%
_{A_{i}}^{out}=\hat{X}_{A_{i}}^{^{\prime }in}$ and $\hat{P}_{A_{i}}^{out}=%
\hat{P}_{A_{i}}^{^{\prime }in}$. By inserting Eq. (\ref{input}) into Eq. (%
\ref{output}), the final output of lights can be written as
\begin{eqnarray}
\left(
\begin{array}{c}
\hat{X}_{L_{i}}^{^{\prime }out} \\
\hat{P}_{L_{i}}^{^{\prime }out}%
\end{array}%
\right)  &=&-C_{1}\left(
\begin{array}{c}
\hat{X}_{L_{i}}^{in} \\
\hat{P}_{L_{i}}^{in}%
\end{array}%
\right) +C_{2}\left(
\begin{array}{c}
\hat{X}_{L_{i}}^{^{\prime }in} \\
\hat{P}_{L_{i}}^{^{\prime }in}%
\end{array}%
\right)   \notag  \label{9} \\
&&-C_{3}\left(
\begin{array}{c}
\hat{X}_{A_{i}}^{in} \\
\hat{P}_{A_{i}}^{in}%
\end{array}%
\right) ,  \label{coefficient}
\end{eqnarray}%
where $C_{1}=1-e^{-\kappa ^{2}}$, $C_{2}=e^{-\kappa ^{2}/2}$, and $%
C_{3}=e^{-\kappa ^{2}/2}\sqrt{1-e^{-\kappa ^{2}}}$. In Fig. \ref{fig3},
these coefficients are plotted versus the coupling strength $\kappa $. From
the diagram, it is obvious that with increasing $\kappa $ the coefficient $%
C_{1}$ increases, $C_{2}$ decreases, and $C_{3}$ attains a maximum.
When $\kappa $ increases to a certain value, $C_{1}$ converges to 1,
and $C_{2}$ and $C_{3}$ tend to zero. The final canonical operators
of output for the read-out pulses will have the following relations
with those of the input for the write-in pulses:
\begin{equation}
\hat{X}_{L_{i}}^{^{\prime }out}=-\hat{X}_{L_{i}}^{in},\text{ \ }\hat{P}%
_{L_{i}}^{^{\prime }out}=-\hat{P}_{L_{i}}^{in},
\end{equation}%
where the noise from the atomic ensembles ($\hat{X}_{A_{i}}^{in}$ and $\hat{P%
}_{A_{i}}^{in}$) and the pulses of light ($\hat{X}_{L_{i}}^{^{\prime
}in}$ and $\hat{P}_{L_{i}}^{^{\prime }in}$) will disappear. So the
final output state of the pulses also satisfies the definition of
the CV PECS for large $\kappa $.

\section{Fidelity analysis}

In this section, we will analyze the fidelities of the storage and
the retrieval by calculating the variances. For the sake of
simplicity, we assume the squeezing parameters $r$ of the four modes
of the CV PECS to be identical.
With $V(\hat{X}_{L_{i}}^{(0)})=V(\hat{P}_{L_{i}}^{(0)})=1/4$, $V(\hat{X}%
_{A_{i}}^{in})=V(\hat{P}_{A_{i}}^{in})=1/4$, and
$V(\hat{X}_{L_{i}}^{^{\prime }in})=V(\hat{P}_{L_{i}}^{^{\prime
}in})=1/4$, we can give the expressions of
the variances, which can be used to characterize the entanglement \cite%
{Loock03,Su07}. For the CV PECS of light to be stored in the atomic
ensembles, the variances of the linear combinations of the components are
\begin{eqnarray}
V_{1}(\hat{P}_{L_{1}}^{in}-\hat{X}_{L_{2}}^{in}) &=&V_{4}(\hat{P}%
_{L_{4}}^{in}-\hat{X}_{L_{3}}^{in})=\frac{1}{2}e^{-2r},  \notag \\
V_{2}(\hat{P}_{L_{2}}^{in}-\hat{X}_{L_{1}}^{in}-\hat{X}_{L_{3}}^{in})
&=&V_{3}(\hat{P}_{L_{3}}^{in}-\hat{X}_{L_{2}}^{in}-\hat{X}_{L_{4}}^{in})=%
\frac{3}{4}e^{-2r}.  \notag \\
&&  \label{process1}
\end{eqnarray}%
For the state stored in the atomic ensembles, the variances are
\begin{eqnarray}
&V_{1}^{^{\prime }}&(\hat{P}_{A_{1}}^{out}-\hat{X}_{A_{2}}^{out})=V_{4}^{^{%
\prime }}(\hat{P}_{A_{4}}^{out}-\hat{X}_{A_{3}}^{out})  \notag \\
&=&\frac{1}{2}(1-e^{-\kappa ^{2}})e^{-2r}+\frac{1}{2}e^{-\kappa ^{2}},
\notag
\end{eqnarray}%
\begin{eqnarray}
&&V_{2}^{^{\prime }}(\hat{P}_{A_{2}}^{out}-\hat{X}_{A_{1}}^{out}-\hat{X}%
_{A_{3}}^{out})  \notag \\
&=&V_{3}^{^{\prime }}(\hat{P}_{A_{3}}^{out}-\hat{X}_{A_{2}}^{out}-\hat{X}%
_{A_{4}}^{out})  \notag \\
&=&\frac{3}{4}(1-e^{-\kappa ^{2}})e^{-2r}+\frac{3}{4}e^{-\kappa ^{2}}.
\label{process2}
\end{eqnarray}%
In addition, for the final output state retrieved from the atomic ensembles,
the variances have the following expressions:
\begin{eqnarray}
&V_{1}^{^{\prime \prime }}&(\hat{P}_{L_{1}}^{^{\prime }out}-\hat{X}%
_{L_{2}}^{^{\prime }out})=V_{4}^{^{\prime \prime }}(\hat{P}%
_{L_{4}}^{^{\prime }out}-\hat{X}_{L_{3}}^{^{\prime }out})  \notag \\
&=&\frac{1}{2}(1-e^{-\kappa ^{2}})^{2}e^{-2r}+\frac{1}{2}(2-e^{-\kappa
^{2}})e^{-\kappa ^{2}},  \notag
\end{eqnarray}%
\begin{eqnarray}
&&V_{2}^{^{\prime \prime }}(\hat{P}_{L_{2}}^{^{\prime }out}-\hat{X}%
_{L_{1}}^{^{\prime }out}-\hat{X}_{L_{3}}^{^{\prime }out})  \notag \\
&=&V_{3}^{^{\prime \prime }}(\hat{P}_{L_{3}}^{^{\prime }out}-\hat{X}%
_{L_{2}}^{^{\prime }out}-\hat{X}_{L_{4}}^{^{\prime }out})  \notag \\
&=&\frac{3}{4}(1-e^{-\kappa ^{2}})^{2}e^{-2r}+\frac{3}{4}(2-e^{-\kappa
^{2}})e^{-\kappa ^{2}}.  \label{process3}
\end{eqnarray}%
Based on these equations, the plots of Fig. \ref{fig4} analyze the
fidelities of this scenario. They show that when $\kappa =1.5$, the
variances corresponding to Eq. (\ref{process1}) tend to zero in the
limit of infinite squeezing (i.e., $r\rightarrow \infty $). However,
the variance corresponding to Eq. (\ref{process2}) converges to a
smaller constant
whereas Eq. (\ref{process3}) converges to a larger constant when $%
r\rightarrow \infty $. Therefore, the variances increase gradually
(i.e., the entanglement decreases gradually) for Eqs.
(\ref{process1}), (\ref{process2}), and (\ref{process3}). Moreover,
when $\kappa =2.5$, we find that all the variances converge to zero
in the limit of infinite squeezing, and they have almost the same
evolution curves. So, the state stored in atomic ensembles and the
state retrieved from atomic ensembles have the same entanglement
properties with the initial CV PECS for large $\kappa $. Thus, the
fidelities of the storage and subsequent retrieval are high for larger $%
\kappa $ and low for smaller $\kappa $. With the increase of the
coupling strength $\kappa $, the protocol will approach perfection,
which can also be understood easily from Eqs.
(\ref{input})-(\ref{coefficient}).

\section{Discussion}

Because the polarization entanglement has some advantages that the
quadrature entanglement does not have, we investigate CV PECSs in
our scheme. For quadrature entanglement, the CV cluster states can
be generated deterministically through offline squeezing and passive
linear optics \cite{Loock07}; moreover, some other
proposals have also been put forward recently \cite%
{Yukawa08,Su07,Menicucci07}. Currently, for polarization
entanglement, schemes for preparing CV cluster states have not been
proposed. However, there are two ways to generate CV polarization
cluster states. CV
polarization entanglement was first realized by Bowen \emph{et al.} \cite%
{Bowen89}, who reported the experimental transformation of the CV
quadrature entanglement between two optical beams into CV
polarization entanglement. Hence, people can also transform
quadrature cluster states into polarization cluster states, taking
advantage of this method. Very recently, another scheme for
efficiently creating CV polarization-entangled states was
experimentally demonstrated by Dong \emph{et al.} \cite{Dong07}, who
used two polarization-squeezed input states to generate polarization
entanglement directly. Thus, CV PECSs can also be generated based on
the scheme of Dong \emph{et al}.

Our protocol is based on the Larmor precession of the atomic spins
in an external field by which the effect of damping is distributed
among both atomic canonical operators \cite{Muschik06}. This leads
to a symmetry between the atomic canonical operators $\hat{X}_{A}$
and $\hat{P}_{A}$, which is a crucial feature for this proposal. In
addition, the fidelities of our protocol can be enhanced by
increasing the coupling strength $\kappa$. The above analysis shows
that when $\kappa$ increases to 2.5, the results are almost perfect.
In Ref. \cite{Duan00}, a detailed analysis of the coupling strength
$\kappa$ is given where $\kappa \sim 5$ is obtainable by adjusting
the detuning and the loss. Moreover, the authors of Refs.
\cite{Takano08,Sherson06,Takeuchi06} have also discussed the
coupling parameter $\kappa$, which can also be changed by adjusting
other parameters in experiment. So our protocol does not place a
very high demand on the value of the coupling strength $\kappa$ and
is experimentally feasible under realistic technological conditions.

\section{Conclusion}

In conclusion, we have proposed a protocol for storing and
retrieving a CV PECS in macroscopic atomic ensembles. Taking the
quadripartite linear CV cluster state as an example, we have
investigated the realization of this scheme in detail and have
discussed the fidelities by calculating the variances. In fact, for
an arbitrary multipartite CV PECS, storage and retrieval can be
achieved using this protocol. Our scheme is made perfectly by
enhancing the coupling strength and is feasible within the current
experimental technology.

\begin{acknowledgments}
We thank Prof. Jietai Jing and Dr. Yu Wang for helpful discussions
and Dr. Florian Hudelist for carefully reading our manuscript and
for providing detailed comments that have assisted us in improving
it. This work was supported by the National Basic Research Program
of China (973 Program) under Grant No. 2011CB921604; the Key Program
of the National Natural Science Foundation of China under Grant No.
60931002; the National Natural Science Foundation of China under
Grants No. 61073048, No. 10874045, and No. 11004059; the Anhui
Provincial Natural Science Foundation under Grant No. 11040606M16;
the China Postdoctoral Science Foundation under Grant No.
20110490825; the Major Program of the Education Department of Anhui
Province under Grant No. KJ2010ZD08; the Key Program of the
Education Department of Anhui Province under Grant No. KJ2010A287;
the Personal Development Foundation of Anhui Province under Grant
No. 2009Z022; the Fundamental Research Funds for the Central
Universities.
\end{acknowledgments}

\end{document}